\documentclass[%
reprint,
amsmath,amssymb,
aps,
superscriptaddress
]{revtex4-2}
\usepackage{hyperref}
\usepackage{color,soul}
\usepackage{amsmath} % ,amsfonts,amsthm: Math packages for equations
\usepackage{graphicx}% Include figure files
\usepackage{dcolumn}% Align table columns on decimal point
\usepackage{bm}% bold math
\DeclareUnicodeCharacter{2212}{\ensuremath{-}}

\newcommand \MZ [1] {\bgroup\noindent[\textcolor{blue}{\bf{MZ}: #1}]\egroup\ignorespacesafterend}

\begin{document}

	\title{Failure precursors and failure mechanisms in hierarchically patterned paper sheets in tensile and creep loading}
	
	\author{Mahshid Pournajar}
	\email{mahshid.poornajar@fau.de}
    \affiliation{WW8-Materials Simulation, Department of Materials Science, Friedrich-Alexander-Universit\"{a}t Erlangen-N\"{u}rnberg, Dr.-Mack-Str. 77, 90762 F\"{u}rth, Germany}
	\author{Tero M\"akinen}
    \affiliation{Department of Applied Physics, Aalto University, P.O. Box 11100, FI-00076 Aalto, Espoo, Finland}
    \author{Seyyed Ahmad Hosseini}
    \affiliation{WW8-Materials Simulation, Department of Materials Science, Friedrich-Alexander-Universit\"{a}t Erlangen-N\"{u}rnberg, Dr.-Mack-Str. 77, 90762 F\"{u}rth, Germany}
    \author{Paolo Moretti}
    \affiliation{WW8-Materials Simulation, Department of Materials Science, Friedrich-Alexander-Universit\"{a}t Erlangen-N\"{u}rnberg, Dr.-Mack-Str. 77, 90762 F\"{u}rth, Germany}
    \author{Mikko Alava}
    \affiliation{Department of Applied Physics, Aalto University, P.O. Box 11100, FI-00076 Aalto, Espoo, Finland}
	\author{Michael Zaiser}
    \affiliation{WW8-Materials Simulation, Department of Materials Science, Friedrich-Alexander-Universit\"{a}t Erlangen-N\"{u}rnberg, Dr.-Mack-Str. 77, 90762 F\"{u}rth, Germany}
	
	\date{\today}
	
	\begin{abstract}
		Quasi-brittle materials endowed with (statistically) self-similar hierarcical microstructures show distinct failure patterns that deviate from the standard scenario of damage accumulation followed by crack nucleation-and-growth. Here we study the failure of paper sheets with hierarchical slice patterns as well as non-hierarchical and unpatterned reference samples, considering both uncracked samples and samples containing a macroscopic crack. Failure is studied under displacement-controlled tensile loading as well as under creep conditions. Acoustic emission records and surface strain patterns are recorded alongside stress-strain and creep curves. The measurements demonstrate that hierarchical patterning efficiently mitigates against strain localization and crack propagation. In tensile loading, this results in a significantly increased residual strength of cracked samples. Under creep conditions, for a given range of lifetimes hierarchically patterned samples are found to sustain larger creep strains at higher stress levels; their creep curves show unusual behavior characterized by multiple creep rate minima due to the repeated arrest of emergent localization bands. 
	\end{abstract}
	
	%\pacs{64.60.av}
	
	\maketitle
\section{Introduction}

Hierarchically (micro)structured materials contain elements that have structure within themselves in such a manner that the system exhibits a self-similar pattern of geometrically similar features on multiple scales. A material's hierarchical order $n$ can be defined as the number of scale levels with recognized structure. For $n = 0$, the physical properties of the material can be analyzed as those of a  structure-less continuum. Materials with $n = 1$ may be envisaged in terms of the arrangement of atoms or granules into ordered crystal or lattice structures \citep{Alava2006}, or into disordered cohesive assemblies. Examples of two-level material architectures include the geometrical microstructure of cellular metamaterials with regular atomic microstructure of the matrix material and well-defined periodic meta-structure. Many biological materials such as nacre, bivalve shell, bone, and tendon collagen have, instead, multi-level hierarchical microstructures which extend from the nanoscale to the macroscale. Although they are made of brittle constituents, such materials may exhibit remarkable stiffness, good toughness, high strength, and low weight \citep{Sun2012, Jiao2015, Gao2006, Rho1998, Gautieri2011, Sen2011}. By mimicking the hierarchical structures of biological materials, novel approaches may be sought to enhance the mechanical properties of architectured (meta)materials \citep{Oftadeh2014}.

Numerical modeling has been considered in several publications as a tool for designing hierarchical structures and predicting their properties. Simple modelling approaches such as fiber bundle models may be extended to hierarchical materials \citep{Biswas2019}, but cannot adequately represent crack-tip stress concentrations which are an essential aspect in the failure of materials by crack propagation. Spatial stress concentrations and crack propagation can be described by fuse and beam network models. As an example, \citet{Moretti2018} used a random fuse model (RFM) to comparatively study failure behaviour of materials with and without hierarchical microstructure. The failure behaviour of hierarchical materials, as well as the resulting super-rough crack morphology, was found to differ significantly from the behavior of non-hierarchical ones. Hierarchical patterning of interfaces was also studied using fuse network models, and it was shown that hierarchical patterning delays detachment and improves interfacial adhesion \citep{Esfandiary2022}. \citet{Hosseini2021} introduced a hierarchical version of a beam network model (BNM) to simulate hierarchically patterned materials. The findings of these authors show that the failure of hierarchically structured materials is caused by local damage nucleation followed by damage percolation rather than critical crack propagation, a finding which holds for both two-dimensional \citep{Hosseini2021} and three-dimensional \citep{Hosseini2022} structures. In both cases, crack profiles show large deflections as failure arises from the coalescence of widely separated micro-cracks. \citet{Zaiser2022} not only conducted simulations but also experiments on two-dimensional quasi-brittle materials loaded along a single axis. It was demonstrated that hierarchical patterning considerably increased crack propagation resistance both in terms of the peak stress and the work of failure of pre-cracked samples. 

These studies indicate that hierarchically structured materials do not fail by coherent crack propagation even when they consist of brittle elements. In materials without hierarchical structure, the applied stress is concentrated at the crack tip and damage or plasticity localize in a narrow zone around the crack tip, known as the fracture process zone (FPZ) \citep{Driscoll2016}. In quasi-brittle materials, such as paper, microcracks nucleate, grow, and coalesce ahead of the crack tip, creating again a fracture process zone of enhanced damage \citep{Bonamy2011}. Here we demonstrate that hierarchical patterning of the material improves toughness and damage tolerance by reducing stress concentrations and significantly widening the process zone in front of the crack tip. To this end, we study paper sheets with hierarchical gap patterns created by a laser cutter, as well as reference specimens containing gaps of equal orientation and overall length in non-hierarchical random arrangement, and pristine sheets. We perform displacement controlled tensile as well as time dependent creep tests to investigate how hierarchical structure affects the failure mode under different loading conditions. We put edge notches of different length into the samples to induce damage localization, and we monitor damage accumulation by acoustic emission  (AE) recording in conjunction with the digital image correlation (DIC) technique to extract displacement fields and spatial strain distribution. In this manner we quantitatively characterize the distinct differences in failure behavior between hierarchical and non-hierarchical samples.

\section{Experimental procedure}

\subsection{Specimen preparation}

Experiments were carried out on rectangular copy paper sheets of thickness $t=0.1$~mm with a weight of 80~g/m$^2$, which were tested under uniaxial loading conditions with loads acting parallel to two of the sides of the sheets. Following the usual geometry of a tensile testing apparatus, the loading direction is henceforth referred to as vertical direction. We used sheets of dimensions $210\times150\times0.1$ mm$^3$ as well as $105\times75\times0.1$ mm$^3$. For the larger samples, two stripes of 30 mm width at the top and bottom edge of the sample were placed in the clamps; for the smaller samples, the width of these stripes was 15 mm; these stripes were left unpatterned. On the central part of the samples, of dimensions $L \times L \times t$ where $L = 150$ mm for the larger and $L=$ 75 mm for the smaller samples, patterns of load parallel (vertical) cuts were created using a laser cutter (model Sculpfun S9).  

\begin{figure*}
% \centering
\includegraphics[width=0.8\textwidth]{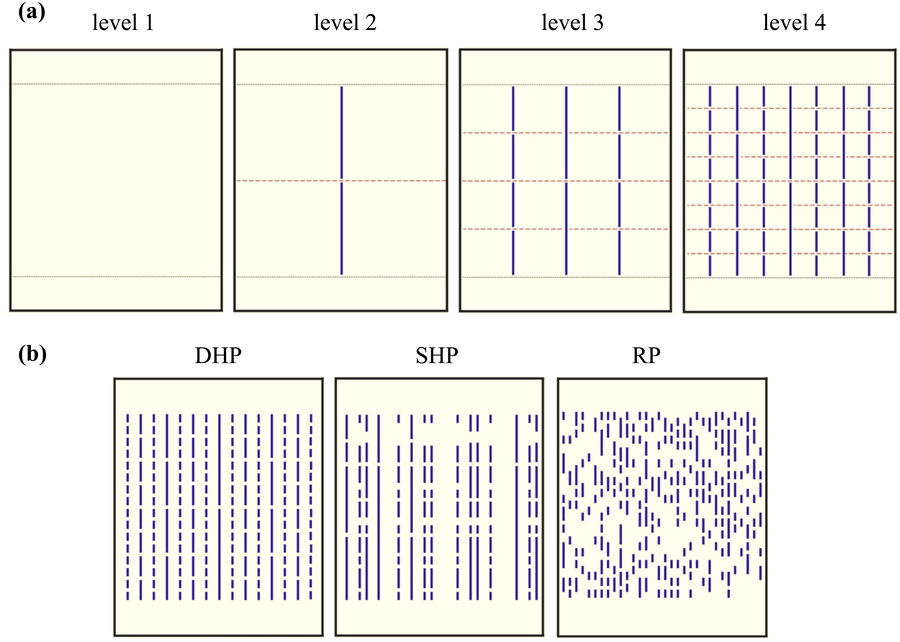}
\caption{(a) Top-down construction of a 4-level deterministic hierarchical pattern (DHP) of cuts: First, we select a paper sheet with dimensions of $210\times150\times0.1$ mm$^3$, allocating stripes of thickness 30 mm at the top and bottom for clamping (level~1); next, the paper is divided into four modules by two load-parallel (vertical) cuts along the centerline of the sample with 1 mm space between them (level 2), here the blue lines are the cuts which have a thickness of 0.1 mm, and the red dashed lines indicate the connectors between them in load-perpendicular direction; next, each of the four modules is again divided by two shorter central cuts into four modules with one module spanning connection (level 3), and after that each of the sixteen modules is again subdivided by two shorter cuts into four modules (level 4); (b) 5-level DHP structure together with structures with shuffled hierarchical pattern (SHP) and random pattern (RP), illustrating the patterns of cuts for the different structures.}
\label{fig:DHP_bottomup_construction}
\end{figure*}

Different types of deterministic and stochastic, hierarchical and non-hierarchical cut patterns were created. We first explain the construction method underlying a deterministic hierarchical pattern (DHP). Top-down construction of a 4-level DHP pattern is illustrated in Fig.~\ref{fig:DHP_bottomup_construction}a. The green dashed lines  indicate the upper and lower edges used for clamping, the space between these lines is denoted as a level-1 load carrying module of size $L$. Next, two central load-parallel cuts are applied to divide this space into four level-2 modules; horizontally a stripe of width $w_{\rm c}=1$ mm between the cuts provides a lateral connection which spans the system in horizontal direction (central dashed line at level 2). The load-parallel cuts are illustrated as blue lines, the red dashed lines illustrate lateral connections between modules. At level 3, each of the four modules is again divided by two central cuts into four level-3 modules, with a horizontal lateral connection of width $w_{\rm c}=1$ mm spanning the modules, and after that, each of the sixteen level-3 modules is again subdivided by two shorter cuts into four modules with one lateral connection to produce a total of 64 level-4 modules. This process is continued for up to $n=7$ levels, the ensuing pattern consists of stripes of paper of width $w_{\rm s} = L/2^{n-1}$ partly separated by cuts. In this pattern, both the lengths of load parallel cuts and the lengths of lateral connections obey power law statistics~\cite{Moretti2018}. 

In addition to the DHP, we also generate a randomized version named shuffled hierarchical pattern (SHP). To this end, we envisage the pattern as a sequence of $2L$ vertical stripes of width $w_{\rm s}/2$ separated by $2L-1$ boundary lines. Along $L$ of these boundary lines, the stripes are fully connected, whereas along the remaining $L-1$ lines, they are connected by discrete lateral connectors of width $w_{\rm c}=1$~mm each, separated by cuts. We now randomly shuffle the boundary lines to create a randomized structure of the same size and total cut length. Next, we envisage the ensuing pattern as a stack of rows. A row of the pattern is a sequence of horizontally adjacent pieces of paper separated by cuts, or a sequence of horizontal connectors separated by cuts. Then we randomly shuffle the rows. As discussed elsewhere \cite{Moretti2018,Hosseini2022}, this randomization leaves the total cut length unchanged. It also preserves the power law statistics of load parallel cuts and of lateral connections, which is constitutive for a hierarchical pattern.

We also generated random patterns (RP) to match the DHP and SHP patterns. A random pattern matching the above described hierarchical patterns again consists of $2L$ continuous stripes of paper of width $w_{\rm s}/2$ divided by sequences of cuts. The number of cuts and the sum of all cut lengths are the same as in the matching hierarchical patterns, as is the minimum spacing $w_{\rm c}=1$mm between two vertically adjacent cuts and the minimum cut length $w_{\rm l}=L/2^{n-1}-w_{\rm c}$. However, now the cut endpoints are located randomly on the boundary lines under the constraints of minimum spacing $w_{\rm c}=1$ mm, minimum cut length, and zero overlap of cuts. This leads to an exponential distribution of cut lengths. A 4-level DHP structure as well as corresponding SHP and RP structures are illustrated in Fig. \ref{fig:DHP_bottomup_construction}b.  
DHP and SHP samples representing 7-level and 6-level structures were used in this work. Paper without any pattern (NP) was also used for comparison. 

In order to investigate flaw sensitivity of failure properties, samples in the as-patterned state were tested alongside with samples containing single load-perpendicular side notches, akin to macroscopic mode-I cracks. Notches with different lengths $a$ ranging from 0 to 0.6 of the specimen width $L$ were cut in load perpendicular direction, starting from the left side of the samples. The notches were vertically located at random positions between 0.25 and 0.75 of the sheet length.

\subsection{Testing and characterization}

An Instron Electropuls E1000 tensile testing machine was used in all tests. Uni-axial displacement controlled tensile tests were done at a displacement rate of 0.05~mm/s. The sample was fixed between two clamps; the upper clamp and grip were movable while the lower one was stationary. All tests were carried out at room temperature. 

On samples of size $210\times150\times0.1$~mm$^3$ with side notches of length $a=30$~mm, also creep tests were performed. 
For such samples, the average peak load in a displacement controlled tensile test was determined to be 265 N for SHP and DHP samples, and 210 N for RP and NP samples. In creep testing, the load was increased at constant rate within 5 s to 85\% of the tensile peak load of the respective type of notched patterned sheet (DHP, SHP, RP, and NP), and then kept constant.  Due to the wide scatter of sample lifetimes, a window of lifetimes ranging from 6 s to 300 s was imposed and samples which did not fail within this time window were discarded. 

Displacement fields were evaluated using a contactless technique called digital image correlation (DIC)~\citep{hild2006digital}. DIC measurements were performed on samples of size $210\times150\times0.1$ mm$^3$ with side notches of length $a=30$~mm. A camera recorded high resolution images (0.25~mm/pixel) of the sample at a rate of one image per second during the total test time from the beginning of the loading until the final failure. Two LED lamps were positioned on both sides of the samples to adjust the light on the surface of the sample. To provide grayscale contrast, a speckle pattern was printed on the specimens. DIC was carried out using Ncorr software \citep{blaber2015ncorr}. In DIC, a reference image is chosen at the beginning of the loading. A region of interest (ROI) is specified on the reference image, which was taken to be an area of $90\times90$~mm$^2$ in front of the crack tip, see the colorscale area in Fig. \ref{fig:DIC_tensiletest}. 
This region is then partitioned into (overlapping) circular subsets, i.e. groups of points over which deformation is assumed to be homogeneous, with centerpoints placed at each pixel in the ROI.
To track the displacement of these subsets DIC tries to find a deformation function that optimally maps the pattern of the reference image onto the pattern of the current test image at time $t$. In our analysis, a subset radius of 16 pixels (corresponding to 4~mm) was used, which was found to provide a reasonable compromise between too-noisy displacement data (too small subset size) and decreased accuracy caused by the smoothing effect of large subsets \citep{Mustalahti2010}. The sum of squared differences (SSD) correlation criterion is used to analyze the similarity between elements in the reference and target images \citep{Pan2009}. The displacement of a point at $(x, y)$ in the reference image is given by a vector $\bm{u} = (u, v)$ and the vertical Green--Lagrange strain used is calculated as
\begin{equation}
    \epsilon_{yy} = \frac{\partial v}{\partial y} + \frac{1}{2} \left[ \left( \frac{\partial u}{\partial y} \right)^2 + \left( \frac{\partial v}{\partial y} \right)^2 \right]
\end{equation}
where the partial derivatives are inferred by locally fitting a plane of radius 3~pixels (0.75~mm) to the displacement field~\citep{blaber2015ncorr}.
\bigbreak
The AE system consisted of a piezoelectric transducer and a rectifying amplifier, A/D-converter, and a computer. The piezoelectric transducer was gently attached directly to the back side of the paper. The transducer was in direct contact with the paper to avoid the requirement for excessive amplification. The AE signal was acquired at an acquisition frequency of 100~kHz during the time of recording stress-strain curves in uni-axial tensile tests and strain-time curves in creep tests. For each type of test and each sample type, AE recordings of 20 samples were taken. 

Acoustic emission signals from deformed paper samples are characterized by an intermittent sequence of discrete AE bursts \cite{Salminen2002}. To analyze statistical signatures of this 'crackling noise'~\citep{sethna2001crackling} signal, the acquired signals were post-processed by transforming them into discrete events using an amplitude threshold $A_{\rm th}$, defining an event as the compact time interval $\Delta t_i$ between sequential upward and downward crossings of the threshold and denoting the time of the upward crossing as the event starting time $t_i$. The threshold is adjusted to separate bursts from background noise, using the fact that for varying $A_{\rm th}$ the total event number $N(A_{\rm th})$ exhibits two distinct regimes, depending whether it is dominated by the discrete bursts or the continuous background noise. 

For each event, the integral of the squared acoustic amplitude $A(t)$ over the event duration was used to calculate the event energy $E_i$,
\begin{equation}
E_i = \int_{\Delta t_i} A^2 (t)dt.
\end{equation}
A set of energy-time pairs $(t_1, E_1), …., (t_n, E_n)$ characterizes the resulting sequence of discrete events. Acoustic event energies $P(E)$ were found to be distribuwaybackted according to a power-law, $P(E) \propto E^{−\beta}$, and the value of the exponent $\beta$ was estimated using the method of maximum-likelihood \citep{baro2012analysis}. This method is based on identifying a $\beta$ value that maximizes the likelihood function. First the original data, here energy values, are restricted between a lower cutoff $E_{\rm low}$ and an upper higher cutoff $E_{\rm up}$. By changing these two values, a map of exponent values can be obtained. By discarding areas where the error estimate of the maximum-likelihood estimate exceeds a prescribed level (here 0.10), an error margin can be set for the exponent (here $\delta \beta = 0.05$). One then identifies the largest area of the map corresponding to $\beta \pm \delta \beta$ and takes this as the estimated $\beta$ value (as well as the corresponding extremal $E_{\rm low}$ and $E_{\rm up}$ values as the limits of the power-law fit)~\citep{reichler2021scalable}.

\section{Results}

\subsection{Displacement controlled tensile test}

\begin{figure*}[p]
%\centering
\includegraphics[width=1\textwidth]{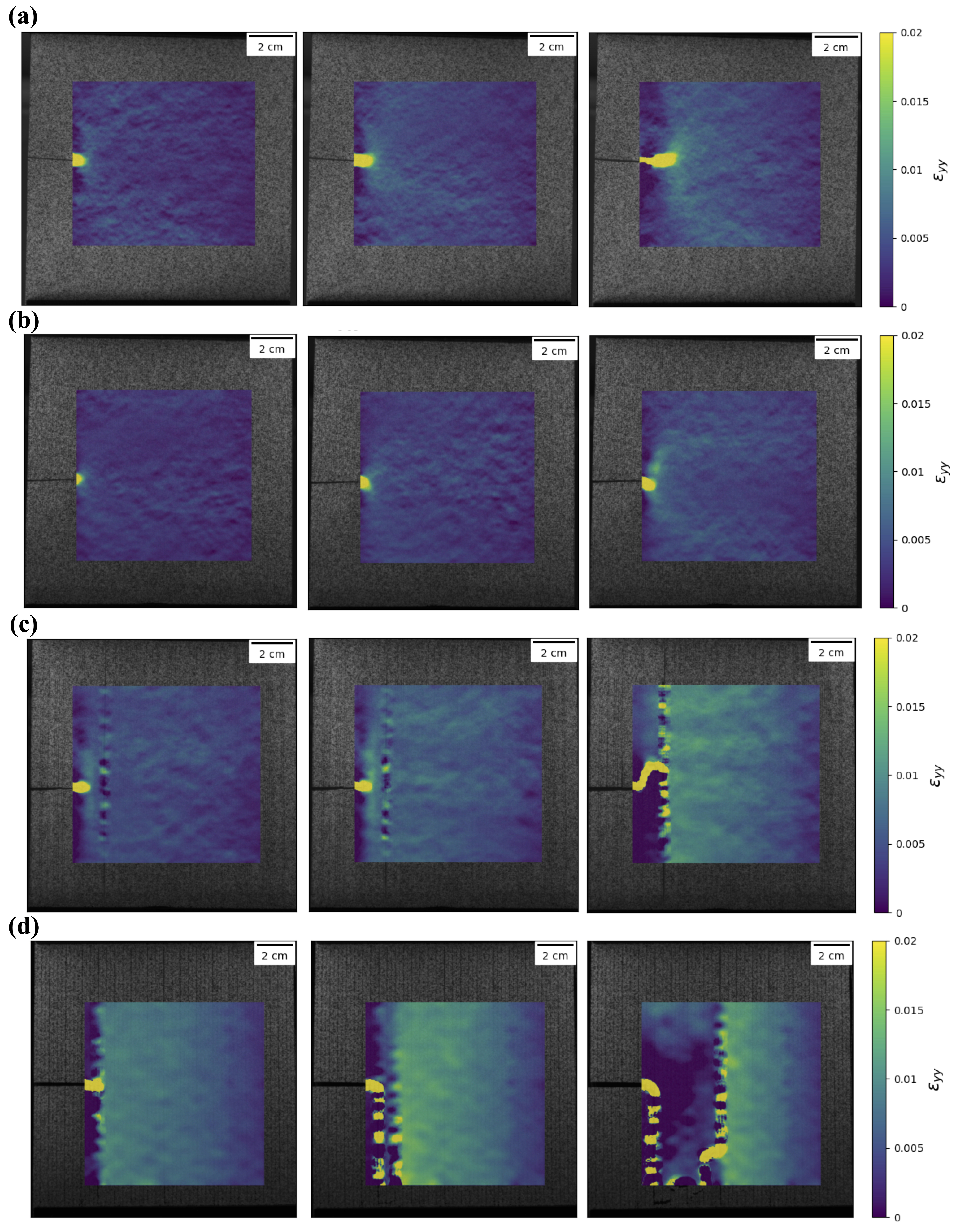}
\caption{DIC maps of the load-parallel strain component $\epsilon_{\rm yy}$ in displacement controlled tensile tests,
displacement level $\varepsilon/\varepsilon_{\rm f} = 0.6$ (left), 0.75 (center), 0.9 (right);  samples with four different cut patterns are shown: a) no pattern (NP),
b) random pattern (RP), c) shuffled hierarchical pattern (SHP), d) deterministic hierarchical pattern (DHP); all samples have a size of 150 mm $\times$ 150 mm with a central side notch of length $a = L/5 = 30$ mm on the left edge; strain is shown in colorscale over a region of interest of 90 mm $\times$ 90 mm, the rest of the samples is shown in greyscale.}
\label{fig:DIC_tensiletest}
\end{figure*}

\begin{figure}[htb]
%\centering
\includegraphics[width=0.5\textwidth]{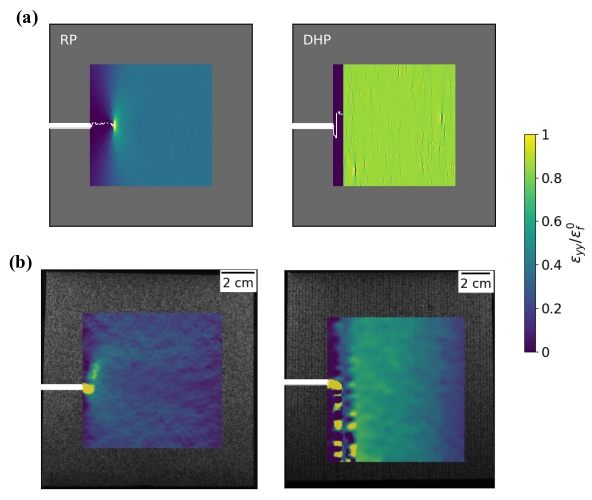}
\caption{Comparison of strain patterns in simulation (a) and experiment (b); left: non-hierarchical (RP) sample, right: hierarchical (DHP) sample; strain patterns are determined  at the peak stress in a displacement controlled test, the initial crack of length a/L = 0.2 is marked in white; the colorscale indicates the local strain, normalized by the mean failure strain $\varepsilon_{\rm f}^0$ of a DHP sample without initial crack.}
\label{fig:simulation}
\end{figure}

We first focus on the effects of patterning on the distribution of strain in tensile testing of notched samples, considering DHP, SHP, RP, and NP samples with a common notch length of 30 mm.  Fig. \ref{fig:DIC_tensiletest} illustrates snapshots of the spatial strain patterns of hierarchical and non-hierarchical samples at 60\%, 75\%, and 90\% of the tensile failure strain $\varepsilon_{\rm f}$ . The color scale in these snapshots indicates the local strain in loading direction. While notched RP and NP specimens exhibit strain concentrations ahead of the crack tip, indicating the presence of a localized fracture process zone, no such strain localization can be seen in hierarchical DHP and SHP samples. At the same time, the overall strain level supported by the hierarchical samples is much higher. The empirical findings were compared with results of simulations based on 2D beam network models (BNM). These simulations consider the same types of structures and the same notch geometry and loading mode as the experiments in Fig. \ref{fig:DIC_tensiletest}. The simulation methodology has been described in detail elsewhere  \citep{Hosseini2021}; we refer the reader to this paper for details. Strain patterns in simulation and experiment are in good qualitative agreement, as shown in Fig. \ref{fig:simulation} for the patterns observed at the peak stress of hierarchical DHP and non-hierarchical RP samples.  The RP samples in both simulation and experiment exhibit a distinct region of high strain around the crack tip, while in DHP samples no discernable crack-tip strain concentration can be found. Accordingly, the crack propagation modes are quite different: in the  non-hierarchical samples, fracture occurs by crack propagation driven by high stresses near the crack tip. In hierarchical samples, on the other hand, the hierarchical gap pattern reduces stress concentrations and impedes crack propagation, leading to early deflection of the crack and diffuse damage accumulation across the entire sample. As a result DHP and SHP samples can absorb significantly more energy and sustain larger deformation before failing. Indeed, in DHP samples the local strain in the intact ligament ahead of the crack may reach levels comparable to the failure strain $\varepsilon_{\rm f}^0$ of an {\em uncracked} sample, demonstrating the efficiency of hierarchical patterns in mitigating the strength deterioration due to the pre-existing crack.

\begin{figure}[htb]
%\centering
\includegraphics[width=0.45\textwidth]{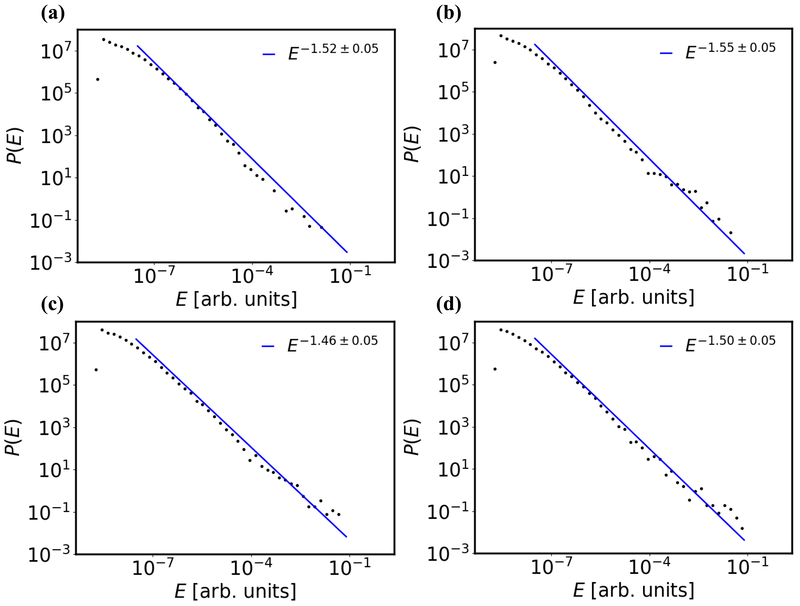}
\caption{Probability density function of event energies as determined from 20 displacement controlled tensile tests. a) NP, b) RP, c) SHP and d) DHP paper sheets. The sample between the clamps has size of 150 mm $\times$ 150 mm with side notch of length $a = L/5 = 30$ mm. The blue lines correspond to the power-law fits given in the figure legends.}
\label{fig:P(E)-tensiletest}
\end{figure}

Acoustic emission allows to monitor the temporal accumulation of damage. Since AE in paper proceeds in the form of a sequence of intermittent events, we consider the statistics of AE event energies and compare results for hierarchical and non-hierarchcial samples. 
Fig.~\ref{fig:P(E)-tensiletest} depicts the probability density function $P(E)$ of event energies for NP, RP, SHP, and DHP samples in displacement controlled tensile tests. The probability density function of acoustic event energies follows a power-law, $P(E) \propto E^{−\beta}$ for around six decades, with $\beta$ values ranging from 1.4 to 1.6. Power law distributions with exponents in this range were observed in samples both with and without an initial notch \citep{Miksic2011b}, e.g., in tensile tests of notched paper, a value of $\beta \sim 1.4\pm0.1$ was determined by Rosti et al. \citep{Rosti2010}. Here we study for the first time avalanche statistics in hierarchically patterned samples and find stable exponent values $\beta \sim 1.5 \pm 0.05$, without statistically significant differences between hierarchical and non-hierarchical samples as shown in Fig. \ref{fig:P(E)-tensiletest}. We note that the upper limits of the maximum likelihood fits coincide with the largest event sizes in the respective statistical samples, indicating that there is no intrinsic upper cut-off to the event size. The size of the largest events is somewhat higher in hierarchical than in non-hierarchical samples. 

The total number of acoustic emission events observed in all 20 experiments during the tensile test is 10287, 5208, 6269, and 5837 for NP, RP, SHP, and DHP, respectively. Hence, 
NP samples seem to exhibit more AE events than patterned samples. This may however be an artefact of the recording method: The cuts in the patterned paper sheets lead to a less perfect contact between the transducer and the paper sheet and at the same time impede acoustic wave transmission, thus reducing reduces the sensitivity of the transducer in recording the sample's released acoustic emission energy.

\subsection{Fracture mechanical size effect}

In fracture mechanics of quasi-brittle materials, the peak stress $\sigma_{\rm p}$ sustained in a tensile test by a sample containing a mode-I crack decreases with increasing crack length $a$. The relationship between peak stress and crack length can, in generalization of Griffith's theory, be written as 
\begin{equation} 
    \label{eq:peak_load_nonhierarchical}
	\sigma_{\rm p} = \dfrac{K_{\rm Ic}}{\sqrt{\pi \left( a + a_{0} \right)}} f \left( \dfrac{a}{L} \right).
\end{equation}
This equation takes into account that in the limit of vanishing crack length, strength must converge to a finite value; the characteristic length $a_0$ can be interpreted as the characteristic size of the fracture process zone (see e.g. \cite{bazant1984size}).  The function $f \left( a/L \right)$ accounts for effects of sample size $L$ and sample geometry, in comparison of geometrically similar samples (fixed $a/L$) this function reduces to a constant. For hierarchical samples, an alternative relationship has been proposed \cite{Zaiser2022} based on the idea that, in such samples, stress concentrations are completely absent such that the strength of a notched sample is simply the strength $\sigma_{\rm p,0}$ of an un-notched sample, corrected by the cross-section reduction due to the initial notch:   
\begin{equation} \label{eq:peak_load_hierarchical}
	\sigma_{\rm p} = \sigma_{\rm p,0} \left( 1 - \dfrac{a}{L} \right). 
\end{equation}

\begin{figure}[thb]
%\centering
\includegraphics[width=0.49\textwidth]{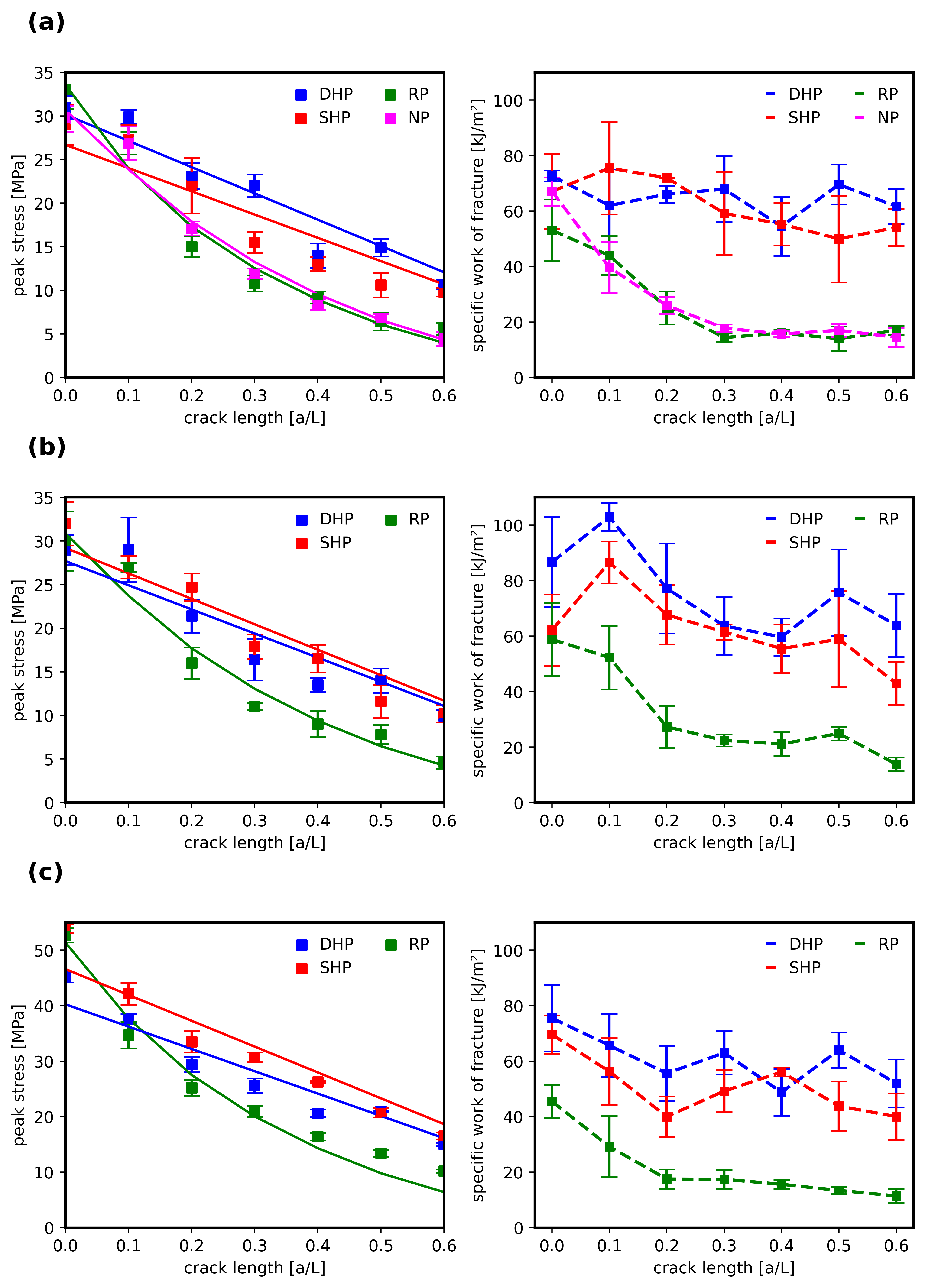}
\caption{Left: Peak stress as a function of notch length, right: specific work of fracture as a function of notch length for NP, RP, SHP and DHP paper sheets in displacement controlled tensile tests; sample size $L$ and number of hierarchy levels $n$ are varied from top to bottom: a) $L = 150$ mm, $n=7$; b) $L = 150$ mm, $n=6$; c) $L=75$ mm, $n=6$.}
\label{fig:strength_crack}
\end{figure}

To investigate how the alternative size effect descriptions apply to hierarchical and non-hierarchical samples, we have plotted the peak stress of square samples containing a side notch as a function of notch length $a$. Different combinations of sample size $L$ ($L = 75$ mm and $L = 150$ mm) and number of hierarchy levels $n$ ($n = 6,7$) were considered, using both NP, RP, DHP and SHP samples; RP samples were matched to the hierarchical samples by using a similar value of the spacing $w_{\rm s} = L/2^{n-1}$ between vertical cuts. Results are shown in Fig. \ref{fig:strength_crack} (left). 
\begin{table}[bt]
  \centering
  \hypertarget{tab:example}{}
  \caption{Summary of fitting parameters for different samples using equations 3 and 4}
  \label{tab:example}
  \includegraphics[width=0.5\textwidth]{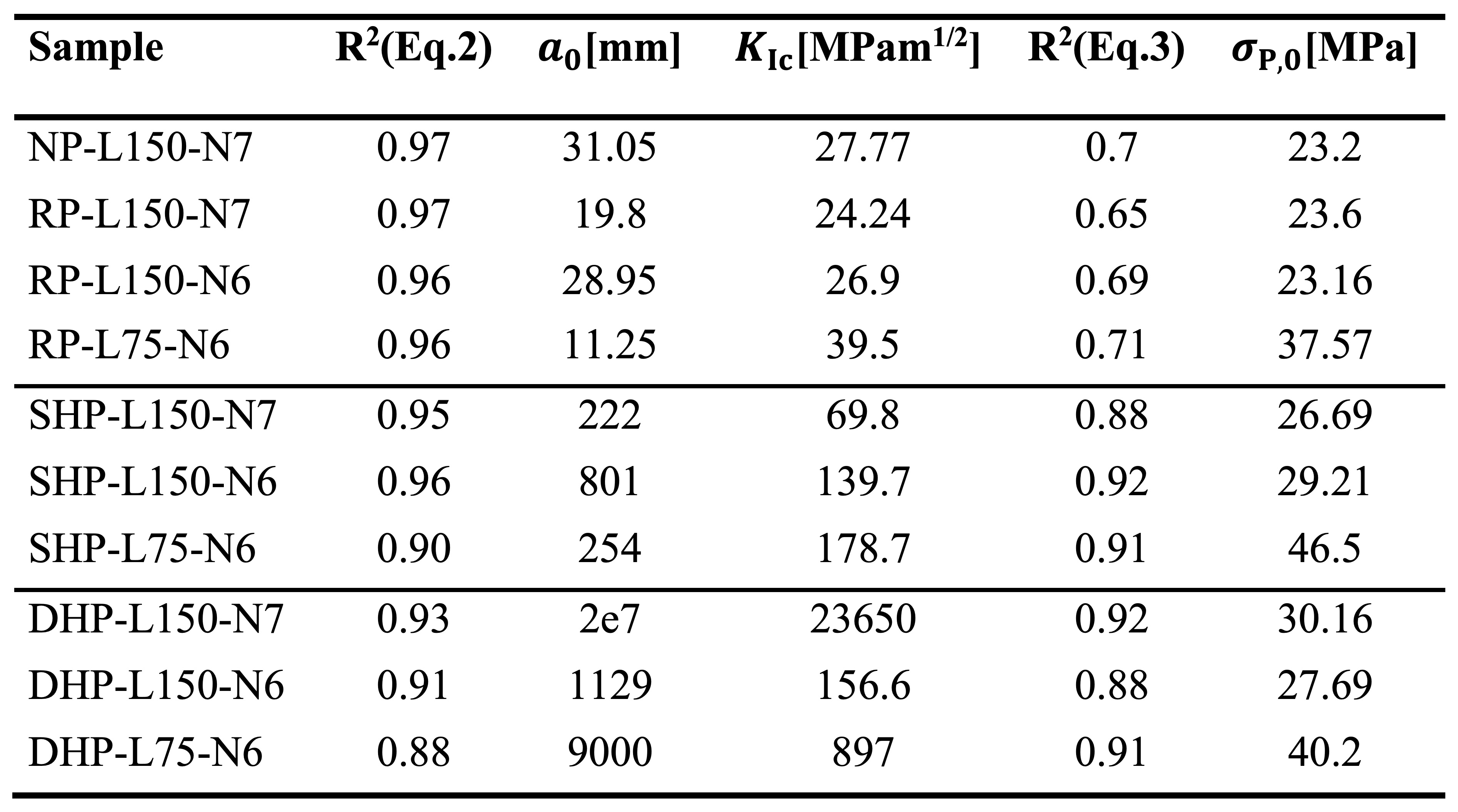}
\end{table}
The data were fitted by both equations \ref{eq:peak_load_nonhierarchical} and \ref{eq:peak_load_hierarchical}, where for $f$ a relationship given by Tada \cite{tada1973stress} for single edge notch specimen was used: $f^{-1}(x)\approx 1.122+0.231x+10.55x^2-21.710x^3+30.382x^4$. Resulting fit parameters are presented in Table \ref{tab:example} alongside values of the coefficient of determination $R^2$ which serves as a measure of the quality of fit.

For non-hierarchical NP and RP samples, Eq. (\ref{eq:peak_load_nonhierarchical}) yields a consistently better description than Eq. (\ref{eq:peak_load_hierarchical}) whereas for SHP and DHP samples, Eqs. (\ref{eq:peak_load_nonhierarchical}) and (\ref{eq:peak_load_hierarchical}) provide a comparable quality of fit. Upon closer inspection of the fit parameters, it is however evident that the parameters obtained for hierarchical samples with Eq. (\ref{eq:peak_load_nonhierarchical}) are unphysical: In all cases, the fitted values of the process zone size exceed the sample size, which implies that a description in terms of fracture mechanics is unfeasible and the obtained fracture toughness values are physically meaningless. We note that fracture testing standards such as ASTM E399 impose the requirement that all relevant dimensions -- specimen size, ligament size, crack length -- must be reasonable multiples of the estimated process zone size. This condition is for the hierarchical samples grossly violated. In our non-hierarchical samples, on the other hand, the condition of a sufficiently small process zone size turns out to be approximately valid. Accordingly, the fit curves shown in \ref{fig:strength_crack} (left) represent fits of Eq. (\ref{eq:peak_load_nonhierarchical}) for non-hierarchical RP and NP samples, and fits of Eq. ( \ref{eq:peak_load_hierarchical}) for hierarchical SHP and DHP samples. 

\begin{figure}[bth]
%\centering
\includegraphics[width=0.4\textwidth]{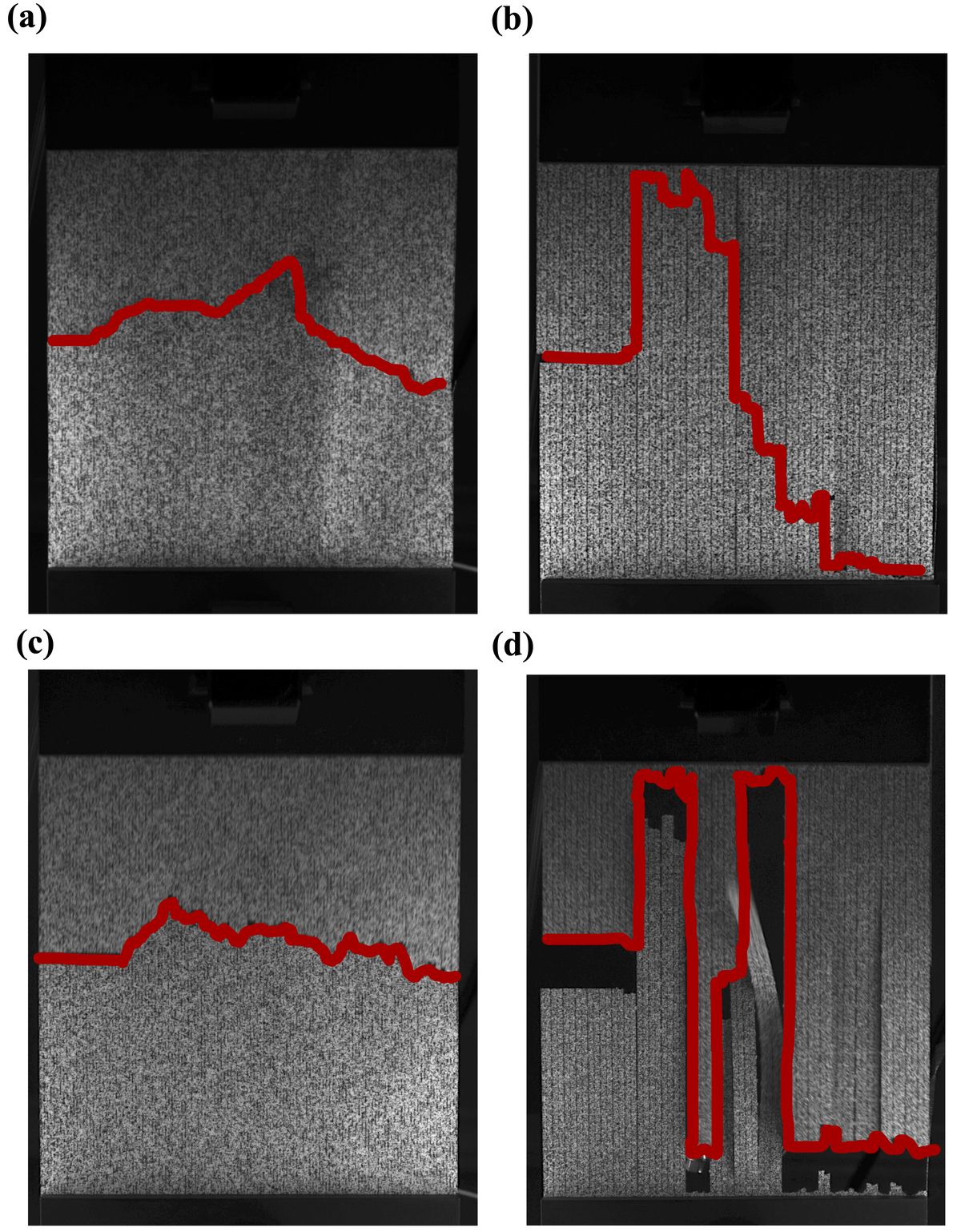}
\caption{Notched paper sheets after fracture in displacement controlled tensile test, a) RP, b) DHP; and after creep test, c) RP and d) DHP. Sample size $L=150$ mm, notch length $a=30$ mm.}
\label{fig:crack_path}
\end{figure}
The above qualitative observations hold for all investigated samples irrespective of sample size and number of hierarchy levels. Comparison of Figs. \ref{fig:strength_crack} (a) and (b) also indicates that strength does not depend strongly on the number of hierarchy levels. At the same time, we note that the strength of samples increases with decreasing sample size, as shown by comparing Figs. \ref{fig:strength_crack} (b) and~(c). The failure strength of samples without cracks is around 50 MPa for $L = 75$ mm and 30 MPa for $L = 150$~mm (see Fig. \ref{fig:strength_crack}). Such sample-size dependence of the strength in un-notched samples is an example of the typical statistical size effects found in disordered materials, where crack nucleation is controlled by the weakest spots of the disordered microstructure -- minima of strength that are, on average, lower in bigger samples \cite{alava2008role}. We also observe that, in samples containing an initial notch, smaller samples show higher strength for the same $a/L$ value. This is entirely in line with the classical fracture mechanical size effect: For a given $a/L$ value, the crack length $a$ in the $L = 75$ mm sample is a factor 2 smaller than in the $L = 150$ mm sample, and following Eq. (\ref{eq:peak_load_nonhierarchical}) the peak stress is accordingly higher. The values of the $a_0$ parameter are in the cases in which Eq. (\ref{eq:peak_load_nonhierarchical}) is applicable quite large compared to the case of ordinary paper, where one would expect $2$ mm.

Fig. \ref{fig:strength_crack} (right) shows the specific work of fracture $w_{\rm f}$, defined as the ratio of the area under the stress-strain curve divided by the area $t(L-a)$ of the intact ligament. In RP samples, regardless of pattern level and specimen size, $w_{\rm f}$ initially decreases with crack length and then stabilizes when the crack length reaches about twice the process zone size. In hierarchical DHP and SHP samples, $w_{\rm f}$ is within the statistical scatter of the data independent of crack length and, for long cracks, a factor of 3-4 higher than in the non-hierarchical RP and NP counterparts. As discussed in \cite{Hosseini2022}, this behavior can be attributed to the fact that the work of fracture is not controlled by the work needed to grow a system spanning crack ('essential work of fracture') but by the work expended to create diffuse damage throughout the sample ('non-essential work of fracture') at locations that do not relate to the pre-existing edge crack. This difference in failure mode is also manifest in the morphology of the fracture paths as depicted in Fig. \ref{fig:crack_path}: 
irrespective of loading mode (displacement controlled tensile or creep test), the hierarchical samples exhibit a super-rough crack surface, which is reminiscent of the simulated fracture patterns of hierarchical structures under tensile testing \citep{Zaiser2022,Hosseini2022} and arises from the coalescence of microcracks disseminated along the entire length of the sample, whose system-spanning coalescence is facilitated by the widely varying length of the power-law distributed vertical cuts. In non-hierarchical NP and RP samples, on the other hand, the fracture surface exhibits a self-affine morphology typical of disordered quasi-brittle materials where failure is governed by crack growth, which is in turn controlled by the interplay of local strength fluctuations and crack-tip stress concentrations \cite{alava2006statistical}. 

\subsection{Creep tests}

\begin{figure}[bt]
%\centering
\includegraphics[width=0.45\textwidth]{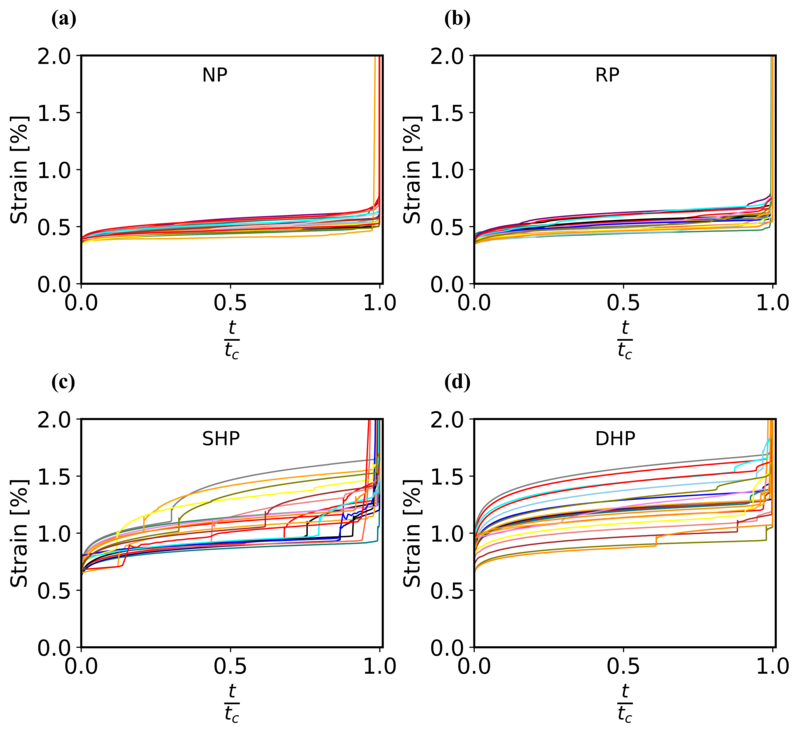}
\caption{Creep strain-time curves under 85\% of peak load for a) NP, and b) RP, c) SHP and d) DHP samples. Each plot shows the results of 20 tested samples of size $L=150$ mm with an initial side notch of length $a = L/5 = 30$ mm.}
\label{fig:creep-strain_time}
\end{figure}

For creep testing we used DHP, SHP, RP, and NP samples of size $150 \times 150$ mm${}^2$ with a 30 mm side notch. These samples were subject to a constant, subcritical stress; the stress level $\sigma_{\rm c}$ was in each case taken to be 85\% of the mean peak stress of the respective sample type in a tensile test. Creep lifetimes show a very significant scatter, the chosen stress level corresponds to typical creep lifetimes $t_{\rm c}$ in the range 5~s~$\le t_{\rm c} \le$~300~s. Samples with creep lifetimes outside this range were discarded for pragmatic reasons.

\begin{figure}[tb]
%\centering
\includegraphics[width=0.5\textwidth]{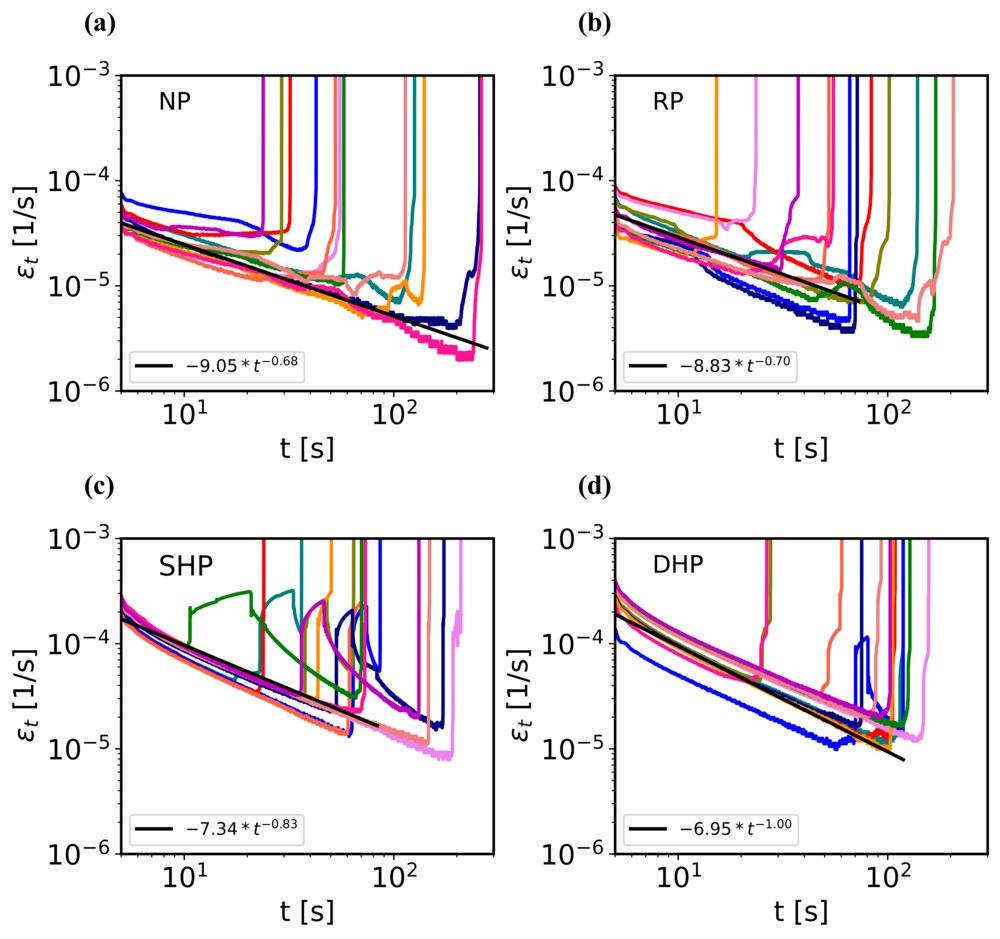}
\caption{Creep rate vs time curves for a) NP, b) RP, c) SHP and d) DHP samples, parameters as in Fig. \ref{fig:creep-strain_time}.}
\label{fig:creep-strainrate_time}
\end{figure}

Fig. \ref{fig:creep-strain_time} shows the creep strain versus the normalized creep time $t/t_{\rm c}$ for DHP, SHP, RP, and NP samples. The first 5~s of a test define the stress ramping time needed to impose the stress $\sigma_{\rm c}$. The ramping time is not shown in the plots, the strain accrued during this time appears as an instantaneous creep strain which defines the initial strain level at the onset of the creep curve. The DHP and SHP structures deform under higher stress; they show an approximately three times higher instantaneous strain as well as a higher overall creep strain as compared to the NP and RP samples, as shown in Fig. \ref{fig:creep-strain_time}.
NP and RP samples exhibit typical three-stage creep curves characterized by a decelerating primary creep stage (stage I) followed by a stage of approximately constant creep rate (stage II) and an accelerating creep stage in the run-up to failure (stage III). The distribution of the creep times or their histogram is relatively wide in all cases. The statistics we have at hand do not allow to make any definite conclusions, but a qualitative comparison with the data of Ref.~\cite{Koivisto2016} and the rather exponential-like life time distribution found there can be made recalling the different experimental parameters (load ratio, sample size) leading to a larger average lifetime. The three hierarchical cases resemble each other, but all also exhibit a few samples with short lifetimes. whereas the NP case has less variation. This difference results as we discuss next from varying mechanisms for crack growth resistance.

\begin{figure*}
%\centering
\includegraphics[width=0.8\textwidth]{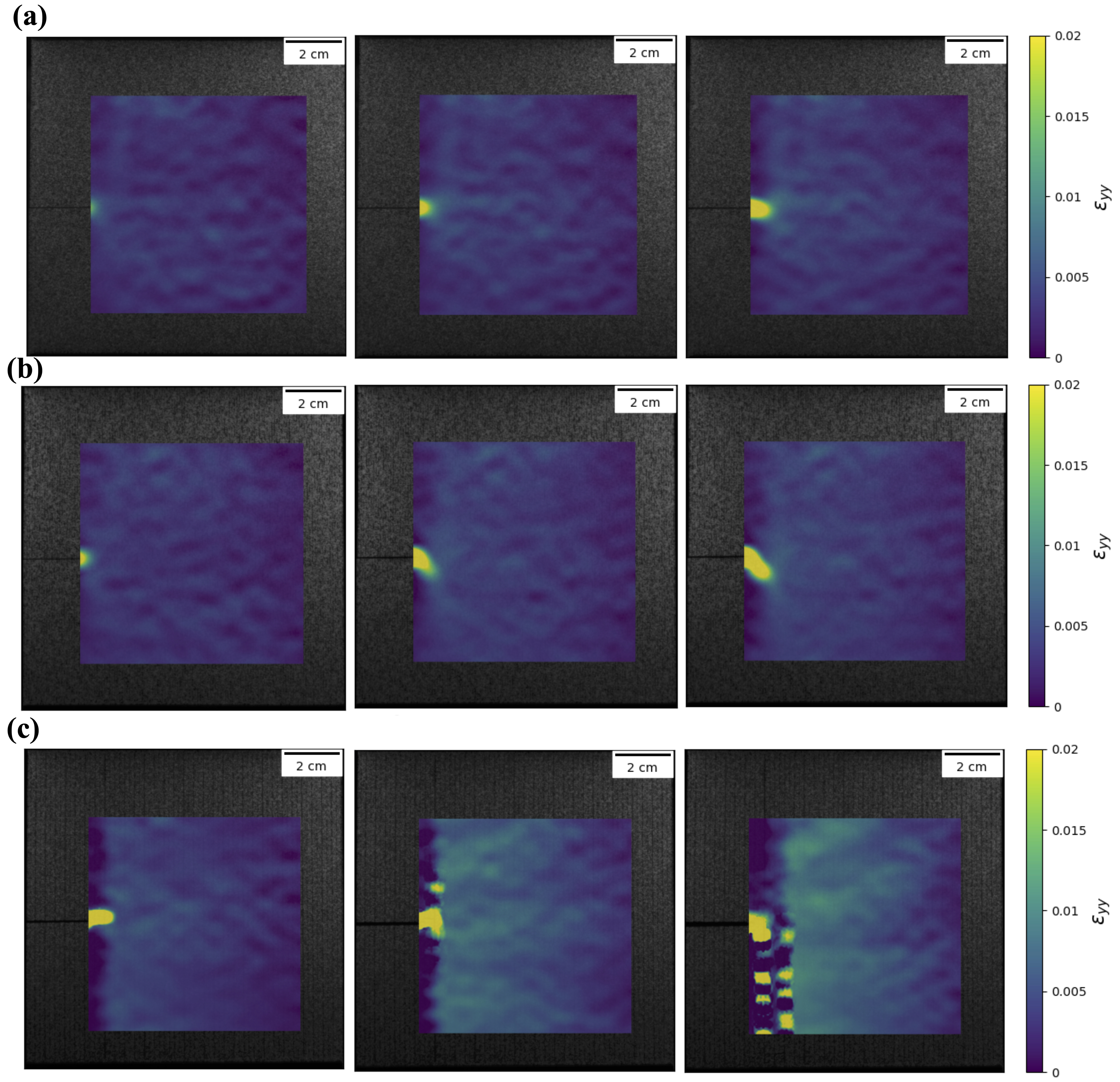}
\caption{Axial strain $\epsilon_{\rm yy}$ map, (left) after ramping, (center) during steady state step, (right) during creep rate acceleration before final failure in a creep test; a) NP, and b) RP and c) DHP samples, sample geometry as in Fig. \ref{fig:creep-strain_time}; the colorscale region of interest is of size 90 mm $\times$ 90 mm.}
\label{fig:DIC_creep}
\end{figure*}

By contrast, hierarchical structures, specifically the SHP, show multi-stage creep curves characterized by an extended stage of primary creep interspersed with sudden strain bursts. This is most evident in Fig. \ref{fig:creep-strainrate_time} showing the time evolution of the creep strain rate of  DHP, SHP, RP, and NP samples after the initial ramping. In the primary creep stage, the strain rate $\epsilon_t$ decreases according to a power law typical of so-called Andrade creep, $\epsilon_t$ $\propto$ $t^{−\alpha}$. The Andrade exponent depends on the type of structure: Whereas non-hierarchical structures exhibit $\alpha$ values close to the classical value of 2/3 ($\alpha = 0.68$ for NP and 0.7 for RP structures), which is consistent with previously reported results for paper\cite{rosti2010fluctuations}, hierarchical structures show higher exponents ($\alpha = 0.83$ for SHP and 1 for DHP structures). The secondary creep stage in non hierarchical samples corresponds to a plateau of the strain rate or to a broad, sometimes fluctuating strain rate minimum. In hierarchical samples, on the other hand, the power-law decrease of the strain rate continues until global strain rate acceleration leads to final failure \citep{Louchet2009, Nechad2005}, but this decrease may be interspersed with sudden strain rate bursts followed by more rapid strain rate relaxation. In view of the fracture surface morphology, which shows the same features as in tensile testing (Fig. \ref{fig:crack_path}) and from comparison with simulations (see Ref. \cite{Hosseini2021}), we may conjecture that this behavior may be associated with the arrest of the initial crack and activation of new cracks at weak spots elsewhere in the sample. This viewpoint is supported by local strain maps obtained by DIC analysis of images taken during the creep test (see Fig. \ref{fig:DIC_creep}). While NP and RP samples show strong strain localization in a pronounced FPZ around the crack tip, the DHP sample shows a much higher overall strain level which is approximately homogeneous across the entire sample and corresponds to a high level of stress everywhere, with a concomitant propensity to distributed damage nucleation.

\begin{figure}[thb]
%\centering
\includegraphics[width=0.5\textwidth]{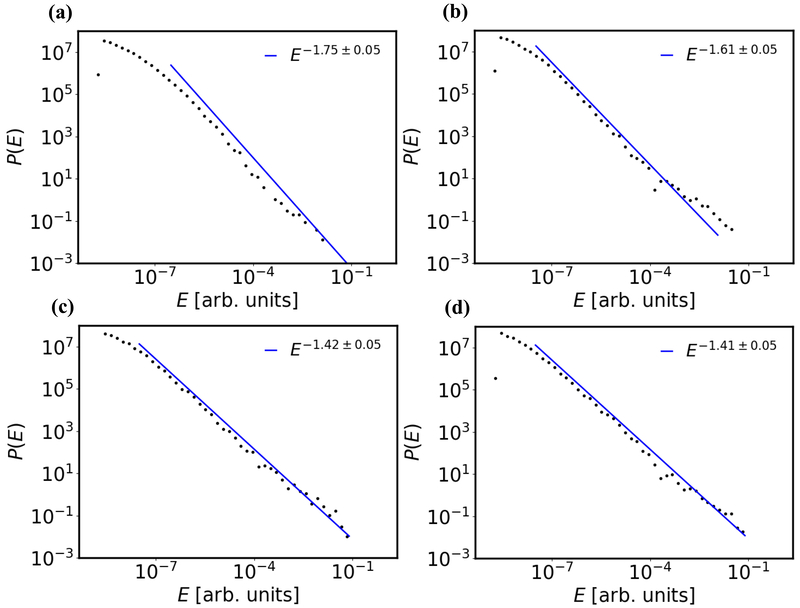}
\caption{Probability density function of event energies for 20 experiments in creep test. a)~NP, b)~RP, c)~SHP and d)~DHP paper sheets. The sample between the clamps has size of $150\times150$~mm${}^2$ with side notch of $a = L/5 = 30$~mm.}
\label{fig:P(E)-creep}
\end{figure}
AE measurements were taken during the creep tests to monitor the time evolution of creep damage. As in tensile testing, the AE signals consist of an intermittent sequence of discrete events (AE bursts). Histograms  $P(E)$ of the acoustic event energies are shown in Fig. \ref{fig:P(E)-creep} for DHP, SHP, RP, and NP samples. Again, we observe for around six decades power law behavior where $P(E) \propto E^{−\beta}$ with exponents $\beta$ that turn out to be similar to the characteristic exponents of AE energy distributions reported in the literature, which were found to lie in the range $1.2< \beta <1.8$ in creep tests of pristine paper \citep{Salminen2002, Viitanen2019}. Maximum likelihood fits result in exponents $\beta = 1.75, 1.61, 1.42$, and 1.41 for NP, RP, SHP, and DHP structures, respectively, with no apparent cut-off at large event energies. Hence, the AE exponents are slightly lower in creep testing of hierarchical structures, as are the total event numbers which amount to $N_{\rm ev}$ = 18534, 8150, 4429, and 4518 for NP, RP, SHP and DHP structures, respectively. Again the sizes of the largest events are higher in the hierarchical structures, reflecting the higher importance of crack nucleation as opposed to incremental crack growth. 

\section{Discussion and Conclusions}

We have conducted a comprehensive investigation of the failure behavior of hierarchically and non hierarchically patterned paper sheets, using DIC and AE in conjunction with tensile and creep testing to determine the spatial and temporal characteristics of damage accumulation in the run-up to failure. Consistent with previous work on tensile deformation which comprises simulation studies \cite{Hosseini2021,Hosseini2022} and experiments on paper and polystyrene sheets \cite{Zaiser2022}, we find significant differences between the failure modes of hierarchical and non-hierarchical samples which are found to hold irrespective of testing mode (displacement controlled tensile vs. creep testing), sample size, and number of hierarchy levels. 

Failure of non hierarchical samples is governed by the paradigm of quasi-brittle fracture mechanics; failure of such samples occurs by crack nucleation-and-propagation or by propagation of a pre-existing crack. Fracture is driven by crack-tip stress concentrations which lead to strain localization in a fracture process zone around the crack tip. As a consequence, the crack size dependence of strength can be well described by fracture mechanics relations with corrections for finite process zone size. 

Failure of hierarchical samples, on the other hand, defies a description in terms of classical fracture mechanics. Attempts to describe the dependency of residual strength on crack length in terms of fracture mechanics lead to internal contradiction as the mathematical fitting procedure results in process zone sizes that exceed the physical size of the sample, in line with the conjecture that structural hierarchy provides a means of shifting the process zone size across several inter-nested hierarchical levels from the elementary to the sample scale \cite{gao2006application,yao2007multi}. On the other hand, a surprisingly good description of the phenomenology is obtained with the naive assumption that, in hierarchical samples, stress/strain concentrations are absent or at least irrelevant. This idea is supported by the findings of simulations as well as by DIC imaging which shows little or no correlation between strain pattern and crack tip location in hierarchical samples. 

While there are distinct differences in failure mode, there are also similarities between hierarchical and non hierarchical samples. In both cases, AE signals are composed of a series of discrete events with power-law distribution of the AE event energies. In both cases, no cut-off can be discerned at large event sizes, and there are only weak differences in the energy exponents although there is a tendency for hierarchical samples to show a lower total event number but larger values for the largest AE event energies. The insensitivity of the AE energy exponent to details of sample morphology and testing mode may be explained by the fact that the observed AE energy exponents $\beta$ are close to the value $\beta = 1.5$ which is the exponent of the critical branching process, hence the mean-field exponent for avalanche processes. This may be taken as one example of universal behavior which, while of academic interest, tends to be irrelevant in the context of materials science: what is universal cannot be changed, hence not be subject of targeted design. What is non universal, on the other hand, can be exploited -- such as the artificial creation of structural hierarchy by targeted patterning which, as our findings demonstrate, may provide an efficient means of enhancíng flaw tolerance of quasi-brittle materials. 

\begin{acknowledgments}
M.P., S.A.H. and M.Z acknowledge support by DFG under grant No. Za171/9-3.  
M.~A. and T.~M. acknowledge support from the Academy of Finland (Center of Excellence program, 278367 and 317464),
FinnCERES flagship (151830423), Business Finland (211835 and 211909), and Future Makers programs.
\end{acknowledgments}

%\bibliography{hierarchicalEXP.bib}
%apsrev4-2.bst 2019-01-14 (MD) hand-edited version of apsrev4-1.bst
%Control: key (0)
%Control: author (8) initials jnrlst
%Control: editor formatted (1) identically to author
%Control: production of article title (0) allowed
%Control: page (0) single
%Control: year (1) truncated
%Control: production of eprint (0) enabled
%

\end{document}